# Preventing corona effects: multi-phosphonic acid poly(ethylene glycol) copolymers for stable stealth iron oxide nanoparticles

**V. Torrisi[1,2], A. Graillot[3], L. Vitorazi[1], Q. Crouzet[3], G. Marletta[2], C. Loubat[3], and J.-F. Berret[1]\***

[1]*Matière et Systèmes Complexes, UMR 7057 CNRS Université Denis Diderot Paris-VII, Bâtiment Condorcet, 10 rue Alice Domon et Léonie Duquet, 75205 Paris, France.*
[2]*Laboratory for Molecular Surface and Nanotechnology (LAMSUN), Department of Chemical Sciences, University of Catania and CSGI, Viale A. Doria 6, 95125, Catania, Italy*
[3]*Specific Polymers, ZAC Via Domitia, 150 Avenue des Cocardières, 34160 Castries, France*

**Abstract:** When disperse in biological fluids, engineered nanoparticles are selectively coated with proteins, resulting in the formation of a protein corona. It is suggested that the protein corona is critical in regulating the conditions of entry into the cytoplasm of living cells. Recent reports describe this phenomenon as ubiquitous and independent of the nature of the particle. For nanomedicine applications however, there is a need to design advanced and cost-effective coatings that are resistant to protein adsorption and that increase the biodistribution *in vivo*. In this study, phosphonic acid poly(ethylene glycol) copolymers were synthesized and used to coat iron oxide particles. The copolymer composition was optimized to provide simple and scalable protocols as well as long-term stability in culture media. It is shown that polymers with multiple phosphonic acid functionalities and PEG chains outperform other types of coating, including ligands, polyelectrolytes and carboxylic acid functionalized PEG. PEGylated particles exhibit moreover exceptional low cellular uptake, of the order of 100 femtograms of iron per cell. The present approach demonstrates that the surface chemistry of engineered particles is a key parameter in the interactions with cells. It also opens up new avenues for the efficient functionalization of inorganic surfaces.

Corresponding author: jean-francois.berret@univ-paris-diderot.fr

This version Thursday, July 24, 2014

# I - Introduction

The concept of "stealth particles" was introduced some years ago to describe therapeutic drug nanocarriers showing an increased blood circulation *in vivo*. The expression was first associated with a series of self-assembled organic particles including liposomes, lipid-based complexes, and biodegradable polymeric micelles.[1] As for the micelles, poly(lactic acid)-*b*-poly(ethylene glycol) or poly(caprolactone)-*b*-poly(ethylene glycol) were among the most studied copolymers because their core-shell structure was found to be resistant to plasma protein adsorption.[2,3] This remarkable property was attributed to the presence of a repulsive poly(ethylene glycol) (PEG) brush playing the role of protective layer, and was further investigated in various drug delivery contexts.[4,5] Poly(ethylene glycol) offers many advantages, among which to be hydrophilic and soluble at body temperature, inexpensive and approved by regulatory health and control agencies.[6]

In parallel, engineered particles with dimensions from 1 to 100 nm made from carbon (nanotubes, graphene) or from metallic atoms (gold, silver, oxide, semi-conductor) were also the subject of intense research during these last years. When disperse in biological fluids however, the particles were found to be selectively coated with serum proteins, which eventually leads to their agglomeration and precipitation.[7-9] In a broad survey on nanoparticle dispersions prior to *in vitro* exposure, Murdock *et al.* demonstrated that many metal (silver, copper) and metal oxide (alumina, silica, anatase) nanomaterials were agglomerating in solutions and that depending on the system, on the presence of serum or on sonication, agglomeration was either agitated or mitigated.[10] The agglomeration of engineered particles is a phenomenon of critical importance since it results in the loss of the nanometer character of the probes, in changes of their hydrodynamic properties and of their interactions with cells.[8,10-13] This behavior was found to be quite general and it has led researchers to propose the paradigm of the "protein corona".[7,8,14-16] Recent reports described this phenomenon as ubiquitous and independent of the nature of particles.[16] The conclusions relative to the protein corona contradict the results found on PEGylated endosomes and micelles, which exhibit a clear and significant resistance against protein adsorption and stealthiness *in vivo*.[1-5] In this work, we address the question of the coating of engineered particles and show that thanks to an appropriate choice of polymer functionalities, stealth iron oxide particles can be obtained.

For applications in nanomedicine, particles need to be functionalized by chemical and physic-chemical modifications of their surfaces. Besides covalent grafting,[17,18] an efficient way to stabilize particles makes use of the layer-by-layer assembly technique.[19] The technique is founded on simple mixing protocols in which charged polymers adsorbed spontaneously on the oppositely charged surfaces by multiple point attachment. Allowing the deposition of single or multi-layers, the technique was applied to various particle types,[20-24] and notably to iron oxide nanocrystals.[20,25-29] In the search of PEGylated coating with better anchoring properties, new polymer architectures have been also proposed. Na *et al.* designed multidentate catechol and PEG derivatized oligomers that provided greatly enhanced stability of iron oxide nanocrystals over a broad range of pH and of electrolytes.[30] To replace the carboxylic groups used as linkage to cationic surfaces, single[31-33] or multiple[34-38] phosphonic acid based polymers were synthesized and tested. Indeed, phosphonic acid has been shown to exhibit a higher binding affinity towards metallic atoms as compared to carboxylic acid, especially in acidic conditions.[36,37,39] Recently, Sandiford *et al.* exploited PEG polymer conjugates containing a terminal bisphosphonic acid group for binding to magnetic nanoparticles. This strategy provided high densities of tethered PEG chains (about 1 PEG per $nm^2$) as well as stable dispersions.[35] Last but not least, phosphonic acid based polymers are recognized for their excellent biocompatibility and for their usefulness in nanomedicine.[37]

In the present study, statistical copolymers containing phosphonic acid and PEG functional groups (Specific Polymers®, France, http://www.specificpolymers.fr/) were synthesized by free radical polymerization. The copolymer composition was optimized to provide a simple and scalable protocol for the preparation of large quantities of products, and an excellent stability of the coating in biological conditions. The efficacy of the phosphonic acid PEG copolymers as a coat was evaluated using iron oxide nanoparticles. Iron oxide is already present in numerous biomedical applications, such as magnetic resonance imaging and hyperthermia, and there is a strong demand for advanced and high-performance coating in these fields. In 2008, our group proposed a simple protocol for the coating of cationic nanoparticles with monofunctionalized phosphonic acid terminated oligomers.[32,33] The same strategy is used here with copolymers that have multiple functional phosphonic acid end-groups. It is demonstrated that the coated particles are stable in biological fluids for months, and that the PEG layer strongly reduces the uptake by living cells, including macrophages.

# II – Experimental Section

*Polymer synthesis*: All phosphonic and carboxylic acid PEG copolymers (patent FR14/00899 [40]) were synthesized by Specific Polymers®, (France, *http://www.specificpolymers.fr/*).

Methacrylic acid (MAA) (CAS: 79-41-4) was supplied by Acros Organics and used as received. PEG-methacrylate (PEGMA, SP-43-3-002, CAS: 26915-72-0) and dimethyl(methacryoyloxy)methyl phosphonate (MAPC1, SP-41-003, CAS: 86242-61-7) monomers were produced by Specific Polymers®. 2,2'-Azobisisobutyronitrile (AIBN) (CAS: 78-67-1) was supplied by Sigma-Aldrich and used after recrystallization in methanol.

Poly(poly(ethylene glycol) methacrylate-co-methacrylic acid), or poly(PEGMA-*co*-MAA) was synthesized by free radical polymerization involving PEGMA and MAA monomers, AIBN as radical initiator and tetrahydrofurane (THF) as solvent. Typical polymerization procedure is described here: MAA (0.21 g, 2.4 mmol), PEGMA (5 g, 2.4 mmol), AIBN (0.02g, 0.12 mmol) were added along with 40 mL THF in a Schlenk flask. The mixture was degassed by three freeze–evacuate–thaw cycles and then heated at 70 °C under argon in a thermostated oil bath for 24 hours, leading to 100% conversion. The copolymers were precipitated in cold ether after synthesis to remove the low molecular weight chains. Poly(PEGMA-*co*-MAA) statistical copolymer was finally recovered after evaporation of the solvents under reduced pressure (Supplementary Information S1).

$^1$H NMR (CDCl$_3$, 300MHz) δ (ppm): 4.0-3.4 (C$\underline{H}_2$-C$\underline{H}_2$-O), 2.2-1.7 (C(CH$_3$)-C$\underline{H}_2$), 0.9-1.5 (C(C$\underline{H}_3$)-CH$_2$).

Poly(poly(ethylene glycol) methacrylate-co-dimethyl(methacryoyloxy)methyl phosphonic acid), or poly(PEGMA-*co*-MAPC1) was synthesized by free radical polymerization involving PEGMA and MAPC1 monomers, AIBN as radical initiator and methylethylketone (MEK) as solvent [40]. Typical polymerization procedure is described here: MAPC1 (0.50 g, 2.4 mmol), PEGMA (5 g, 2.4 mmol), AIBN (0.02 g, 0.12 mmol) were added along with 20 mL MEK in a Schlenk flask. The mixture was degassed by three freeze–evacuate–thaw cycles and then heated at 70 °C under argon in a thermostated oil bath for 24 hours leading to 100% conversion. The PEGMA and MAPC1 monomer conversion rates were measured by $^1$H NMR spectroscopy and were found to have similar time dependences, indicating that the chains have a similar composition throughout the synthesis (S2). The copolymers were precipitated in cold ether after synthesis to remove the low molecular weight chains. Poly(PEGMA-*co*-MAPC1) statistical copolymer was recovered after evaporation of the solvents under reduce pressure.

[1]H NMR (CDCl$_3$, 300MHz) δ (ppm): 4.4-3.5 (C$\underline{H}_2$-C$\underline{H}_2$-O, O-C$\underline{H}_2$-P, O=P-(OC$\underline{H}_3$)$_2$), 2.1-1.7 (C(CH$_3$)-C$\underline{H}_2$), 0.9-1.3 (C(C$\underline{H}_3$)-CH$_2$). [31]P NMR (CDCl$_3$, 300MHz) δ (ppm): 21.8 ppm

The hydrolysis of the phosphonated ester into phosphonic acid was performed using bromotrimethylsilane and then methanol at room temperature as already reported in the literature.[41,42] Solvents were finally evaporated under reduce pressure leading to the Poly(PEGMA-*co*-MAPC1$_{acid}$) final product. The scheme of the synthesis of phosphonic acid poly(ethylene glycol) copolymers is shown in Fig. 1[40].

[1]H NMR (CDCl$_3$, 300MHz) δ (ppm): 4.5-3.5 (C$\underline{H}_2$-C$\underline{H}_2$-O, O-C$\underline{H}_2$-P), 2.1-1.7 (C(CH$_3$)-C$\underline{H}_2$), 0.9-1.3 (C(C$\underline{H}_3$)-CH$_2$). [31]P NMR (CDCl$_3$, 300MHz) δ (ppm): 18.0 ppm

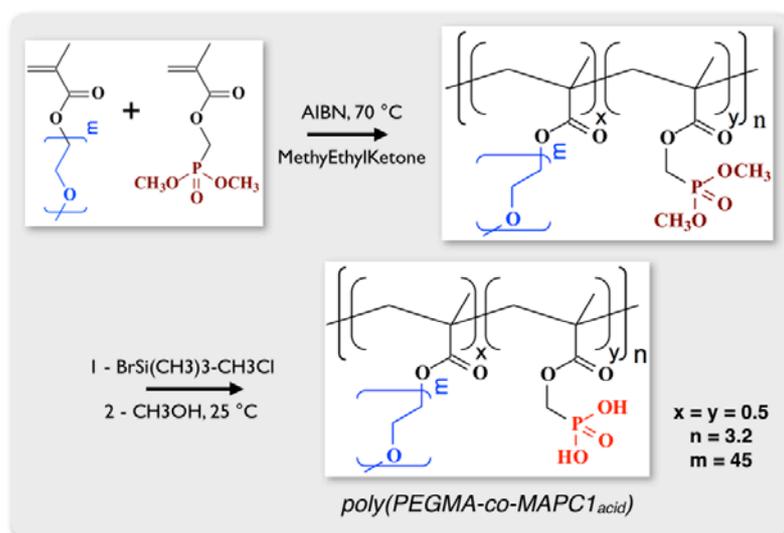

**Figure 1**: *Two-step synthesis of poly(poly(ethylene glycol) methacrylate-co-dimethyl(methacryoyloxy) methyl phosphonic acid), abbreviated as poly(PEGMA-co-MAPC1$_{acid}$) and described as phosphonic acid PEG copolymer in this work.*

*Polymer Characterization:* [1]H NMR and [31]P NMR (δ, ppm): [1]H and [31]P NMR spectra were recorded using a Bruker Avance 300 (300 MHz) with CDCl$_3$ as solvent. For [1]H NMR, chemical shifts were referenced to the peak of residual non-deuterated solvents at 7.26 ppm for CDCl$_3$.

The weight-averaged molecular weight of the polymers was determined by static light scattering measurements using a NanoZS Zetasizer from Malvern Instrument. Quartz cuvettes compatible with water solutions and toluene for calibration were used. The polymers solutions were prepared with 18.2 MΩ MilliQ water, filtered with 0.2 μm cellulose filters and their *pH* was adjusted to 8 by addition of ammonium hydroxide. The Rayleigh ratios $\mathcal{R}(c)$ was measured as a function of the concentration. It was found to vary linearly with $c$ between 0 and 1 wt. % (Supplementary Information S3). The molecular weight of the polymer was derived from the Zimm representation. In this representation, the intercept of the ratio $Kc/\mathcal{R}(c)$ at zero concentration provides the inverse molecular weight $1/M_w$ (S3). Here $K$ is the scattering constrast, which depends on the refractive index increment $dn/dc$ of the polymer solution. The refractive index increment was determined by refractometry (Cordouan Technologies) (S4). For phosphonic and carboxylic acid PEG copolymers, we obtained $dn/dc = 0.148 \pm 0.06 \ cm^3 g^{-1}$ and $dn/dc = 0.137 \pm 0.07 \ cm^3 g^{-1}$, respectively. These values are close to that of PEG, $dn/dc = 0.135 \ cm^3 g^{-1}$, indicating that the contrast is dominated by that of the PEG-segments.

The refractive index increment $dn/dc$, the scattering contrast $K$ and molecular weight of the two polymers are listed in Table I. The $M_w^{Pol}$-values were 12950 $\pm$ 500 g mol[-1] and 15500 $\pm$ 2000 g mol[-1] for the polymers with phosphonic and carboxylic acid anchoring groups, in good agreement with those targeted by the synthesis (11000 g mol[-1]).

The molar-mass dispersity for poly(PEGMA-*co*-MAPC1) and poly(PEGMA-*co*-MAA) were determined by size exclusion chromatography using polystyrene columns and found at 1.81 and 1.78 respectively (see Supplementary Information S2 for details). From the molar equivalent of acid groups per gram (as determined by [1]H NMR, Table I), the average number of functional moieties per chain was estimated. It was found to be n = 3.1 $\pm$ 0.2 for copolymers with phosphonic diacid moieties and n = 4.0 $\pm$ 0.5 for polymers carrying the carboxylic acid groups. From the copolymer molecular weights, and considering the different repetitive units ($M_n(PEGMA) = 2069$ g mol$^{-1}$; $M_n(MAPC1_{acid}) = 208$ g mol$^{-1}$; $M_n(MAA) = 86$ g mol$^{-1}$), the average number of functional groups was found at 3.1 $\pm$ 0.2 and 4.0 $\pm$ 0.5 for polymers carrying phosphonic and carboxylic end-groups, respectively. Both determinations agree with each other, and confirm that the copolymers have multiple functional groups.

| polymer name | acid group meq/g | $dn/dc$ $cm^3 g^{-1}$ | $K$ $cm^2 g^{-2}$ | $M_w^{Pol} g \ mol^{-1}$ | functional group per chain |
|---|---|---|---|---|---|
| phosphonic acid PEG copolymer | 0.873 | 0.148 | $1.60 \times 10^{-7}$ | 12950 $\pm$ 500 | 3.1 $\pm$ 0.2 |
| carboxylic acid PEG copolymer | 0.460 | 0.137 | $1.37 \times 10^{-7}$ | 15500 $\pm$ 2000 | 4.0 $\pm$ 0.5 |

**Table I:** *Structural parameters of the PEG copolymers used in this work. The first column denotes the molar equivalent of acid groups per gram (milli eq $g^{-1}$) of polymer as determined by [1]H NMR. From the refractive index increment $dn/dc$, the scattering contrast $K$ is calculated. The weight-averaged molecular weights were obtained by static light scattering. The numbers of the functional groups are determined from the molecular weight and from the molar equivalent of acid groups. Note that the refractive index increments are close to that of poly(ethylene glycol), $0.135 \ cm^3 g^{-1}$.[32]*

*Nanoparticles:* The iron oxide nanoparticles (bulk mass density $\rho$ = 5100 kg m$^{-3}$) were synthesized according to the Massart method by alkaline co-precipitation of iron(II) and iron(III) salts and oxidation of the magnetite (Fe$_3$O$_4$) into maghemite ($\gamma$-Fe$_2$O$_3$).[20] The nanoparticles were then size-sorted by subsequent phase separations. At $pH$ 2, the particles are positively charged, with nitrate counterions adsorbed on their surfaces. The resulting interparticle interactions are repulsive, and impart an excellent colloidal stability to the dispersion. For the present study, two batches of γ-Fe$_2$O$_3$ nanoparticles of median diameter 6.8 and 13.2 nm were synthesized. The size and size distribution were retrieved from Vibrating Sample Magnetometry (VSM) and from Transmission Electron Microscopy (TEM). Table II lists the results obtained for $D_0^{VSM}$ and $D_0^{TEM}$ and for the size dispersities, $s^{VSM}$ and $s^{TEM}$. For colloids, the size dispersity $s$ is defined as the ratio between standard deviation and average diameter. For the calculation of the number-averaged molecular weight $M_n^{Part}$ in Eq. 1, the molar-mass dispersity of the particles was estimated from the log-normal size distribution obtained by TEM (Fig. S5). Based on these approximations, the 6.8 nm and 13.2 nm particles have a molecular-weight disperstity of 1.43 and 1.61 respectively. The difference between the VSM and TEM sizes originate from noncrystalline defects located close to the particle surface, and not contributing to magnetic properties. VSM was also used to determine the volumetric magnetization $m_S$ of the nanoparticles, which was found to be 2.9×10$^5$ A m$^{-1}$, *i.e.* slightly lower than the volumetric magnetization of bulk maghemite ($m_S$ = 3.9×10$^5$ A m$^{-1}$). From electron microdiffraction scattering, the crystallinity of the particles was demonstrated by the appearance of five diffraction rings which wave vectors matched precisely those of the maghemite structure (S6-S7). In this work, the nanoparticles concentrations are defined by the percentage by weight of γ-Fe$_2$O$_3$ in the dispersion or by the iron molar concentration [Fe]. With these units, $c$(γ-Fe$_2$O$_3$) = 8×10$^{-3}$ wt. % or 80 µg/ml corresponds to [Fe] = 1 mM.

| | 6.8 nm γ-Fe$_2$O$_3$ nanoparticles | 13.2 nm γ-Fe$_2$O$_3$ nanoparticles |
|---|---|---|
| $D_0^{VSM}$ (nm) | 6.7 | 10.7 |
| $s^{VSM}$ | 0.21 | 0.33 |
| $D_0^{TEM}$ (nm) | 6.8 | 13.2 |
| $s^{TEM}$ | 0.18 | 0.23 |
| $M_W^{Part}$ (g mol$^{-1}$) | 1.3×10$^6$ | 12×10$^6$ |
| $D_H$ (nm) | 13 | 27 |

**Table II**: *Characteristics of the iron oxide particles used in this work. $D_0^{VSM}$ and $D_0^{TEM}$ denote the median diameter of the bare particles determined by Vibrating Sample Magnetometry (VSM) and by transmission electron microscopy (TEM). Similarly, $s^{VSM}$ and $s^{TEM}$ are the values of the size dispersity. $D_H$ is the hydrodynamic diameter of the bare particles in water. The weight-averaged molecular weight $M_W^{Part}$ was determined from static light scattering.[43]*

*Coating:* Citric acid is a weak triacid of molecular weight $M_w$ = 192 g mol$^{-1}$ with three acidity constants (pK$_{A1}$ = 3.1, pK$_{A2}$ = 4.8 and pK$_{A3}$ = 6.4). Complexation of the surface charges with citric acid was performed during the synthesis through simple mixing. At pH 8, citrate-coated particles are stabilized by electrostatics. As a ligand, citrate ions were characterized by adsorption isotherms and the adsorbed species were in equilibrium with free citrates in the bulk. The concentration of free citrates was kept at the value of 8 mM,[44] both in DI-water and in culture medium. The hydrodynamic diameter of the citrate-coated particles was identical to that of bare particles, indicating a layer thickness under 1 nm (Table III). Poly(sodium acrylate), the salt form of poly(acrylic acid) (PAA) with a weight-average molecular weight $M_w$ = 2000 and 5000 g mol$^{-1}$ and a polydispersity of 1.7 was purchased from Sigma Aldrich and used without purification. To adsorb polyelectrolytes on the particles, the precipitation-redispersion protocol was applied.[20,45] The precipitation of the iron oxide dispersion by PAA was performed in acidic conditions (*pH* 2). The precipitate was then separated by magnetic sedimentation and its pH was increased by addition of ammonium hydroxide. The precipitate redispersed spontaneously at *pH* 8. The hydrodynamic sizes of PAA$_{2/5K}$ coated γ-Fe$_2$O$_3$ are listed in Table III. These values were 5 and 10 nm larger than the hydrodynamic diameter of the uncoated particles, indicating a corona thickness 2.5 nm and 5 nm respectively. Values of the electrophoretic mobilities and zeta-potentials are provided in the Supporting Informations section (S8).

|  | Hydrodynamic diameter $D_H$ (nm) | |
|---|---|---|
| **coating** | 6.8 nm γ-Fe$_2$O$_3$ nanoparticles | 13.2 nm γ-Fe$_2$O$_3$ nanoparticles |
| citrate | 13 ± 1 | 27 ± 2 |
| poly(acrylic acid) 2K | 18 ± 2 | 32 ± 2 |
| poly(acrylic acid) 5K | 23 ± 2 | 37 ± 3 |
| carboxylic acid-PEG | 26 ± 2 | 50 ± 4 |
| phosphonic acid-PEG | 25 ± 2 | 37 ± 2 |

***Table III***: *hydrodynamic diameter $D_H$ of coated particles as determined by dynamic light scattering.*

*Cell Culture:* Adherent cells from mice including NIH/3T3 fibroblasts and RAW264.7 macrophages were studied. Fibroblasts are the most common cells of connective tissues in animals, in particular in the skin whereas macrophages are professional phagocytes which role is to engulf and eliminate cellular debris and circulating pathogens. NIH/3T3 fibroblast cells were grown in T25-flasks as a monolayer in Dulbecco's Modified Eagle's Medium (DMEM) with high glucose (4.5 g L$^{-1}$) and stable glutamine (PAA Laboratories GmbH, Austria) (S9). The medium was supplemented with 10% fetal bovine serum (FBS) and 1% penicillin/streptomycin (PAA Laboratories GmbH, Austria). Exponentially growing cultures were maintained in a humidified atmosphere of 5% CO$_2$ and 95% air at 37°C, and in these conditions the plating efficiency was 70 − 90% and the cell duplication time was 12 − 14 h. Cell cultures were passaged twice weekly using trypsin–EDTA (PAA Laboratories GmbH, Austria) to detach the cells from their culture flasks and wells. The cells were pelleted by centrifugation at 1200 rpm for 5 min. Supernatants were removed and cell pellets were re-suspended in assay medium and counted using a Malassez counting chamber.

The RAW264.7 were grown in suspension in T25-flasks in Roswell Park Memorial Institute (RPMI) with high glucose (2.0 g L$^{-1}$) and stable glutamine (PAA Laboratories GmbH, Austria). RPMI was supplemented with HEPES 10 mM, 10% FBS and 1% penicillin/streptomycin. The culture and counting protocols for the cells in suspension were similar to those of the NIH/3T3 fibroblasts. Note finally that the macrophages have a duplication time similar to that of fibroblasts.

*Static and dynamic light scattering:* static light scattering is a non-invasive technique used to characterize polymers and particles in solution. In this experiment, the dispersion prepared at a concentration $c$ was illuminated by a laser source. The scattered light intensity $I_S(c, \theta)$ was recorded at a given angle $\theta$ over a period of 5 mn and averaged. The molecular weight of the scatterers was determined by measuring samples of different concentrations and by calculating the Rayleigh ratio $\mathcal{R}(c, \theta)$:

$$\mathcal{R}(c, \theta) = \frac{I_S(c, \theta) n_0^2}{I_T n_T^2} \mathcal{R}_T \tag{1}$$

where $I_S(c, \theta)$ and $I_T$ are the scattering intensities of the sample and that of toluene respectively, $n_0$ and $n_T$ their refractive indexes and $\mathcal{R}_T$ the Rayleigh ratio of toluene. In the present study, we had $n_0 = 1.333$, $n_T = 1.497$ and $\mathcal{R}_T = 1.352 \times 10^{-5} \, cm^{-1}$ and $\lambda = 633 \, nm$. For the analysis of the scattering data, the Zimm representation was used. This Zimm representation consists in plotting the ratio $Kc/\mathcal{R}(c, \theta)$ *versus* $c$. For dilute solutions, it varies as:

$$\frac{Kc}{\mathcal{R}(c, \theta)} = \left( \frac{1}{M_w} + 2A_2 c \right) P(\theta) \tag{2}$$

where $P(\theta)$ is the form factor of the scattering particles, $A_2$ the 2nd virial coefficient and $K = \frac{2\pi^2 n_0^2}{\lambda^4 \mathcal{N}_A} \left( \frac{dn}{dc} \right)^2$ ($\mathcal{N}_A$ being the Avogadro number and $\lambda$ is the wave-length of light). In the $Kc/\mathcal{R}(c, \theta)$ *versus* $c$ representation, the inverse molecular weight is the ordinate at origin and slope is twice the virial coefficient. For particles and/or polymers in the nanometer range as it is the case here, $P(\theta) = 1$.[46]

With dynamical light scattering, the collective diffusion coefficient $D(c)$ was determined from the second-order autocorrelation function of the scattered light. The autocorrelation functions were analyzed using the cumulants and the CONTIN algorithm fitting procedure provided by the instrument software, and it was checked that both gave comparable results. The hydrodynamic sizes determined by light scattering $D_H$ were found to be systematically larger than those obtained by electron microscopy ($D_0^{TEM}$) or by magnetometry ($D_0^{VSM}$). The reason for this difference is related to the size dispersity of the particles, light scattering being sensitive to the largest objects of the distribution.

*Mass of Iron Internalized/adsorbed by Living Cells:* The live cells were cultured and incubated with the particles for 24 h. The supernatant was removed and the cells were thoroughly washed with phosphate buffer saline (PBS). The cells were trypsinized, numbered using a Mallasez chamber and centrifuged. The pellets were dissolved in hydrochloric acid (35 %), and later investigated by UV-Vis spectrophotometry. The cell pellets dissolved in HCl displayed the yellow color characteristic of tetrachloroferrate ions $FeCl_4^-$. The absorbance of the dissolved pellets was compared to those of iron oxide and of cells determined separately. For the sake of accuracy, MILC was calibrated against regular titration techniques such as flame atomic absorption spectroscopy.[47] Assuming for the absorbance an absolute uncertainty of 0.03, the minimum amount of iron detectable by this technique is 0.03 picogram (*i.e.* 30 femtogram) per cell. With MILC, the high sensitivity arises from the fact that the whole UV-Vis spectrum was taken into account in the adjustment. For the 13.2 nm particles used in this study, a mass of iron of $m_{Fe} = 1$ pg/cell corresponds to $5 \times 10^4$ particles.

*Toxicity assays:* The method measured the mitochondrial activity of cells. Subconfluent cell cultures (90% confluency at treatment time) on 96 well plates were treated with 100 µL/well of nanoparticles at different concentrations for 24 h (or at different time points according to the final endpoint), culture medium was removed, cells were rinsed with culture media without phenol red and incubated with 100 µL/well of 2-(4-iodophenyl)-3-(4-nitrophenyl)-5-(2,4-disulphophenyl)-2H-tetrazolium (WST-1, Roche Diagnostics), diluted 1/10 (or more according to cell lines) in culture medium without phenol red for 1 to 4 h. The assay principle is based upon the reduction of the tetrazolium salt WST-1 to formazan by cellular dehydrogenases. The generation of the dark yellow colored formazan was measured at 450 nm in a multiwell-plate reader against a blank containing culture media and WST-1 and it was corrected from the absorbance at 630 nm. The optical density of the supernatant is directly correlated to cell number.

# III - Results and Discussion

## III.1 - Phase behavior and PEG coating

Poly(poly(ethylene glycol) methacrylate-*co*-dimethyl(methacryoyloxy)methyl phosphonic acid), abbreviated in the following poly(PEGMA-*co*-MAPC1$_{acid}$) and depicted as phosphonic acid PEG copolymer (Fig. 1) was synthesized and characterized according to techniques provided in the Materials & Methods and in Supplementary Sections (S1-S4). This polymer was compared to a poly(poly(ethylene glycol) methacrylate-*co*-methacrylic acid), in short poly(PEGMA-*co*-MAA), which presented the same composition as poly(PEGMA-*co*-MAPC1$_{acid}$), but where carboxylic groups replace the phosphonic acid moieties. Solutions of iron oxide nanoparticles and polymers at the same concentration ($c = 0.1$ wt. %) and same *pH* (*pH* 2) were mixed at different volume ratios $X$. This method ensures that the total concentration remains constant[48] and that no aggregation of particles occurs during mixing because of *pH* or salinity gap. The *pH* of the mixed solution was then raised to 8 by addition of ammonium hydroxide. Figs. 2a and 2b displays the stability diagram of mixed dispersions at physiological conditions for 6.8 and 13.2 nm iron oxides mixed with phosphonic acid PEG copolymers.

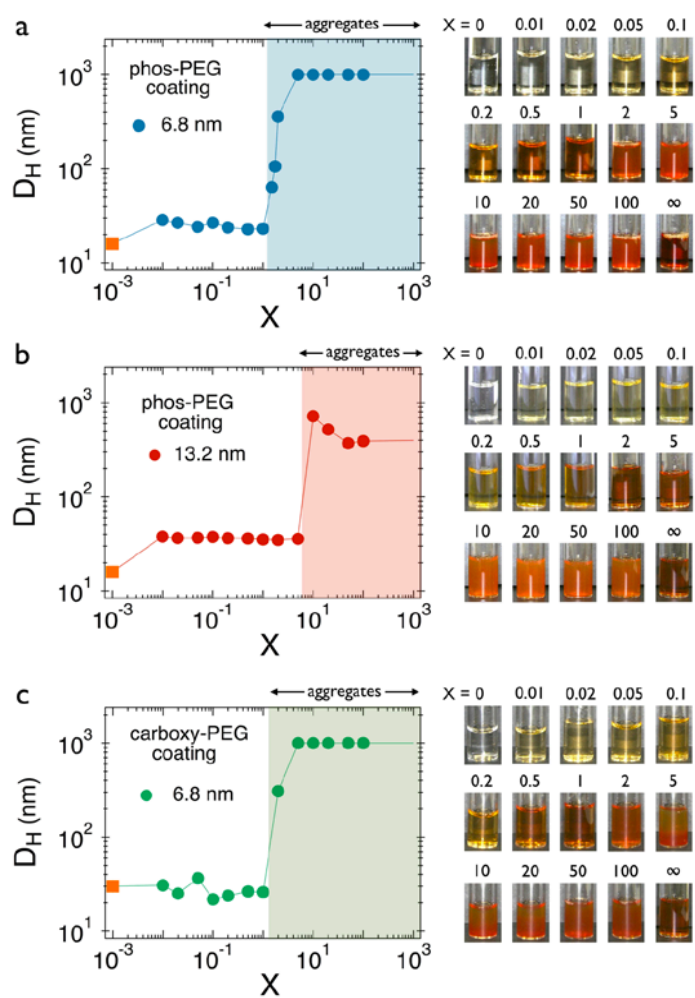

**Figure 2**: *Stability diagram of mixed γ-Fe₂O₃/polymers dispersions at pH8 as a function of the mixing ratio X. X is defined as the ratio between the volumes of nanoparticle and polymer dispersions: a) phosphonic acid PEG copolymers with 6.8 nm particles, b) phosphonic acid PEG copolymers with 13.2 nm particles and c) carboxylic acid PEG copolymers with 6.8 nm particles. The hydrodynamic diameters were measured from samples displayed on the right hand-side. Above a critical ratio $X_C$, dispersions are turbid and the particles agglomerate. For micron-sized aggregates, $D_H$ is set at 1 μm for the sake of simplicity.*

On the left hand side, the hydrodynamic diameter $D_H$ is displayed as a function of $X$. On the right hand-side, images of the vials containing the dispersions are presented. The color change of the dispersions is due to the changes in iron oxide concentration, this concentration varying as $c_{Part} = cX/(1 + X)$. Likewise, the polymer concentration decreased with $X$ as $c_{Pol} = c/(1 + X)$. The hydrodynamic diameter of single polymer chains was 16 nm, as indicated by the square at $X = 10^{-3}$. Upon addition of particles, $D_H$ exhibits first a plateau over 2 to 3 decades in $X$, and then increases sharply with the formation of large micron-size aggregates. The precipitated ranges are identified by colored backgrounds for each assay. In this range, the dispersions are turbid and sediment with time. Supplementary investigations were carried out with the carboxylic acid PEG copolymers and 6.8 nm $\gamma$–$Fe_2O_3$ (Fig. 2c). As for the phosphonic based chains, a broad stability plateau was observed at low mixing ratios. Above the critical value of $X_C = 1.5$, the plateau was followed by a precipitation range. The existence of a critical mixing ratio $X_C$ suggests that adsorption occurred *via* a non-stoichiometric ligand binding process.[32,33] The model assumes that at each value of $X$, the polymers are equally distributed among the particles in the dispersion.[32,48] Below $X_C$, the iron oxide surfaces are saturated with adsorbed polymers, the functional end-groups exceeding the number of binding sites. Above, the coverage is incomplete and the particles precipitate upon $pH$ increase (as uncoated particles do). Here, we exploit this feature to derive the number of adsorbed chains per particle $n_{ads}$. $n_{ads}$ is given by the relation:[48]

$$n_{ads} = \frac{1}{X_C} \frac{M_n^{Part}}{M_n^{Pol}}$$

(1)

where $M_n^{Part}$ and $M_n^{Pol}$ are the molecular weights of the particle and polymer, respectively. For the 6.8 nm iron oxide particles (Figure 2a), $X_C = 1.3$ and $n_{ads} = 97$. For the 13.2 nm particles (Figure 2b), $X_C = 5$ and $n_{ads} = 207$. These $n_{ads}$-values correspond to a polymer density of 0.50 ± 0.15 nm$^{-2}$. A density of adsorbed chains independent of the diameter implies a constant density of structural charges at the iron oxide surface, a result that is expected for this type of particles.[43] A polymer density of 0.50 nm$^{-2}$ corresponds to a coverage of 1.5 PEG chain per nm$^2$, one of the highest densities reported for core-shell assembled structures.[6,49,50] In the plateau regime, the hydrodynamic sizes of the two coated particles were found at 25 and 37 nm respectively, *i.e.* 12 and 10 nm larger than those of the bare particles (Tables III). This result indicates the existence of a 5 - 6 nm thick PEG layer around the iron oxide nanocrystals, consistent with partially stretched PEG chains.[51] With zeta potentials of -6 mV for both particles, electrokinetic measurements confirmed that the PEGylated particles were globally neutral (S8). A schematic representation of the coating is illustrated in Fig. 3.

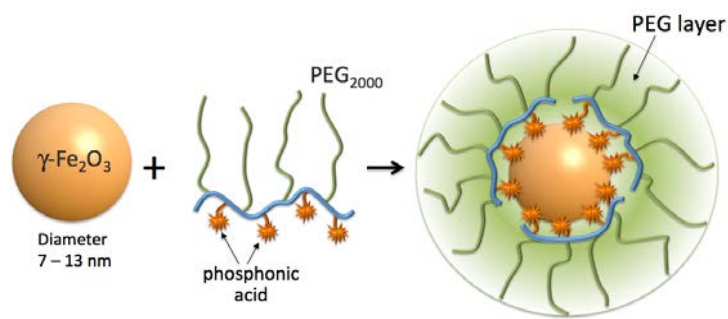

*Figure 3*: *Schematic representation of an iron oxide particle coated with phosphonic acid PEG copolymers.*

## III.2 - Colloidal stability in cell culture media

In the present study, particles were said to be stable if their hydrodynamic size remained constant as a function of the time after a change of medium, and equal to that found in DI-water at physiological $pH$. The stability criterion defined here takes into account the possible scenarios observed in cell culture media, *i.e.* the adsorption of proteins at the outer coating layer, aggregation or both.[10-12,14-16,33,47,52-55] To assess the colloidal stability of coated iron oxides, the following protocol was outlined. A few microliters of a concentrated dispersion were poured and homogenized rapidly in 1 ml of the solvent to be studied, and simultaneously the scattered intensity $I_S$ and diameter $D_H$ were measured by light scattering. The targeted weight concentration was 0.1 wt. %. After mixing, the measurements were monitored over a period of 2 hours. Subsequent measurements of the intensity and diameter were carried out at 1 day, 1 week and 1 month. For particles aggregating over time, both $I_S$ and $D_H$ are expected to increase as compared to their initial values. Iron oxides coated with phosphonic or carboxylic acid PEG copolymers were formulated following the method set out in the Section II.1. The mixing was performed slightly below $X_C$ *i.e.* where the dispersions are stable whatever the $pH$. The dispersions were dialyzed against deionized water using a 50 kD cut-off membrane to remove the excess polymer, and further concentrated by ultrafiltration. Iron oxide with the functional PEG polymers was first studied in buffer (PBS) and in $NH_4Cl$ salted solutions and displayed excellent stability over several weeks (S10). In a second study, cell culture media including Dulbecco's Modified Eagle's Medium (DMEM) and Roswell Park Memorial Institute medium (RPMI), with and without fetal bovine serum were investigated. Fig. 4a displays the hydrodynamic diameter $D_H$ of 13.2 nm iron oxide coated with various polymers. As the concentrated dispersion is added to the cell medium, $D_H$ exhibits a jump from 18 nm to 40-50 nm. The initial value of 18 nm before mixing corresponds to the average size of the proteins and biological macromolecules present in the medium. For 2 hours $D_H(t)$ remains unchanged for particles coated with phosphonic acid PEG copolymers, whereas it increases continuously for particles coated with carboxylic acid PEG copolymers. This later evolution is indicative of a slow destabilization of the dispersion, which is due to the detachment of the coating [55]) or to the slow association with binding macromolecules in the medium. The results on PEGylated particles are compared with those prepared with a poly(acrylic acid) layer. For the 13.2 nm particles, destabilization also occurs, as evidenced by the slow increase of the hydrodynamic diameter as a function of time. The destabilization of $PAA_{2K/5K}$ coated particles is ascribed to the polyelectrolyte properties of the layer. In physiological solvent, the charged polymer brush undergoes a slight collapse due to ionic strength effects, and the thickness of the brush drops below a critical value. This critical value was computed recently for iron oxide nanocrystals by Faure *et al.*[56] and it was found to be 3 nm for 13.2 nm particles. Hence, for particles with a layer thinner than 3 nm, van der Waals forces dominate the particle-particle interactions, and induce agglomeration. This effect is not present for 6 nm thick PEG coating which is insensitive to the ionic strength.

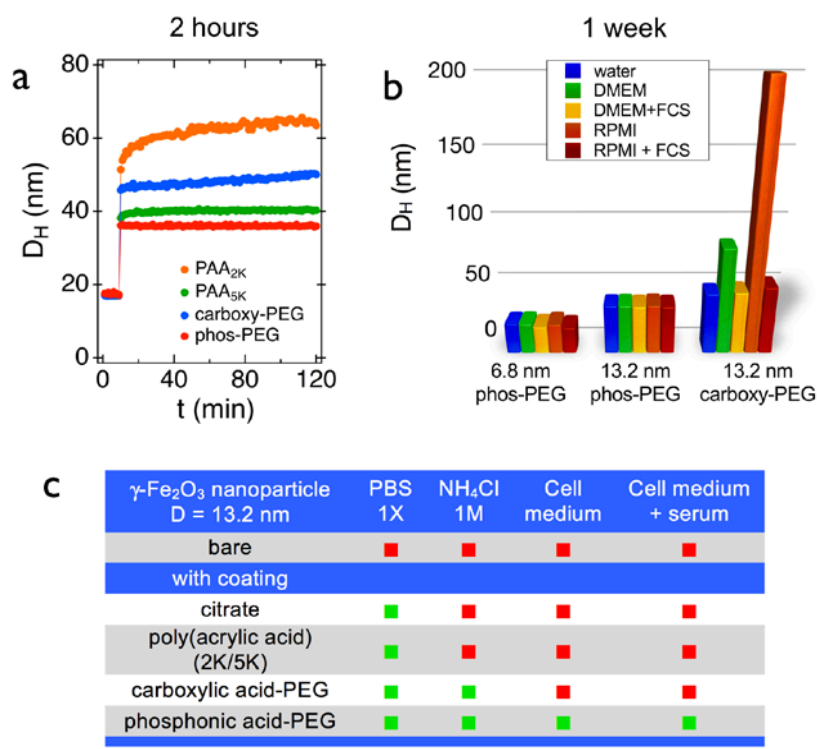

***Figure 4: a)*** *Hydrodynamic diameter* $D_H(t)$ *for 13.2 nm iron oxide suspended in cell medium and coated with poly(acrylic acid) 2100 and 5100 g mol⁻¹ and with phosphonic or carboxylic acid PEG copolymers. Here the solvent was the Dulbecco's Modified Eagle's Medium (DMEM) supplemented with serum. The slow increase in $D_H$ after the initial jump is indicative of the destabilization of the dispersion. Only particles covered with phosphonic acid PEG copolymers are stable and devoid of plasma proteins.* ***b)*** *One-week stability diagram for particles dispersed in water and cell culture media DMEM and RPMI, with and without fetal calf serum (FCS).* ***c)*** *Comparison of the colloidal stability of bare and coated 13.2 nm γ-Fe₂O₃ nanoparticles in various suspending media. Green squares indicate that the particles are stable, red squares unstable.*

After one week, the stability diagram of Fig. 4b exhibited similar features: the 6.8 and 13.2 nm particles covered with phosphonic acid PEG chains were still disperse, whereas partial aggregation was observed with the carboxylic acid PEG polymers. Note that for the carboxylic acid modified polymer, the particles are slightly more stable in the presence of serum, indicating the role played by the proteins in the interactions.[8,57,58] However, over a longer period the carboxylic acid PEG coating leads to a macroscopic aggregation and sedimentation. Fig. 4c summarizes the long-term stability behavior of 13.2 nm iron oxide particles in various solvents, including PBS, NH₄Cl 1M solution and cell culture media. Comparison is extended to bare particles and to particles covered with citrate ligands.[55,59] From these data, it can be seen that polymers with multiple phophonate functionalities and PEG chains outperform all other types of coating examined.

# III.3 - In vitro assays and toxicity

In *in vitro* assays, the amount of particles taken up by the cells is of particular interest for it is a reliable and quantitative evaluation of the interactions with cells. With magnetite or maghemite, this amount is expressed in terms of mass iron atom per cell, noted $m_{Fe}$ and expressed in pg/cell.[47,60,61] Here, we exploit the MILC protocol (*Mass of metal Internalized/Adsorbed by Living Cells*) to measure $m_{Fe}$ in NIH/3T3 mouse fibroblasts and in RAW 264.7 macrophages.[47] Both cell lines are representative of *in vitro* assays currently performed. Based on the digestion of incubated cells (Figs. 5a and 5b) with concentrated hydrochloric acid reactant (Fig. 5c) and complemented by a colorimetric assay (Fig. 5d), the technique detects particles that are adsorbed at the plasma membrane or internalized by the cells. In this assay, the 13.2 nm γ–Fe$_2$O$_3$ particles were selected and coated with 3 different types of ligands or polymers: citrate, poly(acrylic acid) and the phosphonic acid PEG copolymer. MILC was performed in 6-well plates cultivated with 3 million cells per Petri dish in average (Fig. 5). Incubation time was 24 h. The concentrations investigated were $[Fe]$ = 0.03, 0.1, 0.3, 1, 3 and 10 mM, and covered the range of those reported in the literature for *in vivo* and *in vitro* assays.[8,62] The data are displayed in Fig. 6 for the NIH/3T3 and RAW264.7.

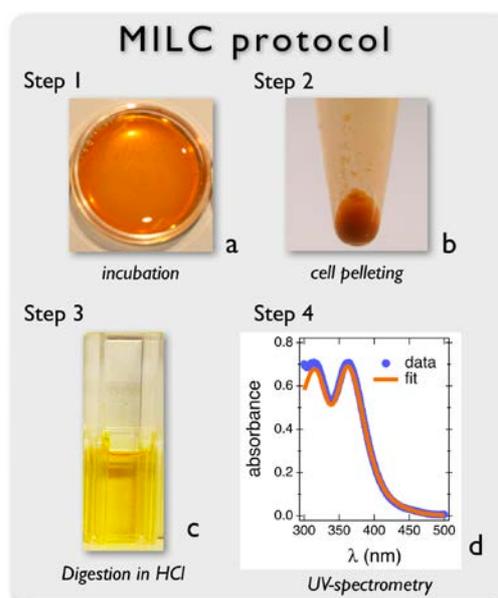

**Figure 5**: *Description of the MILC protocol (**M**ass of metal **I**nternalized/Adsorbed by **L**iving **C**ells) used to measure the mass of iron taken up by living cells.[47] The MILC protocol comprises 4 tasks: a) incubation of living cells with an iron oxide dispersion. b) After 24 h, the cells are pelleted. c) The pellets are digested by a 35 vol. % hydrochloride solution and analyzed by UV-spectrometry. The yellow color arises from tetrachloroferrate ions $FeCl_4^-$. d) Absorbance curves for cells treated with citrate coated particles (closed circles), together with the calibration standard (continuous line) obtained from iron oxide dispersions of known concentrations.*

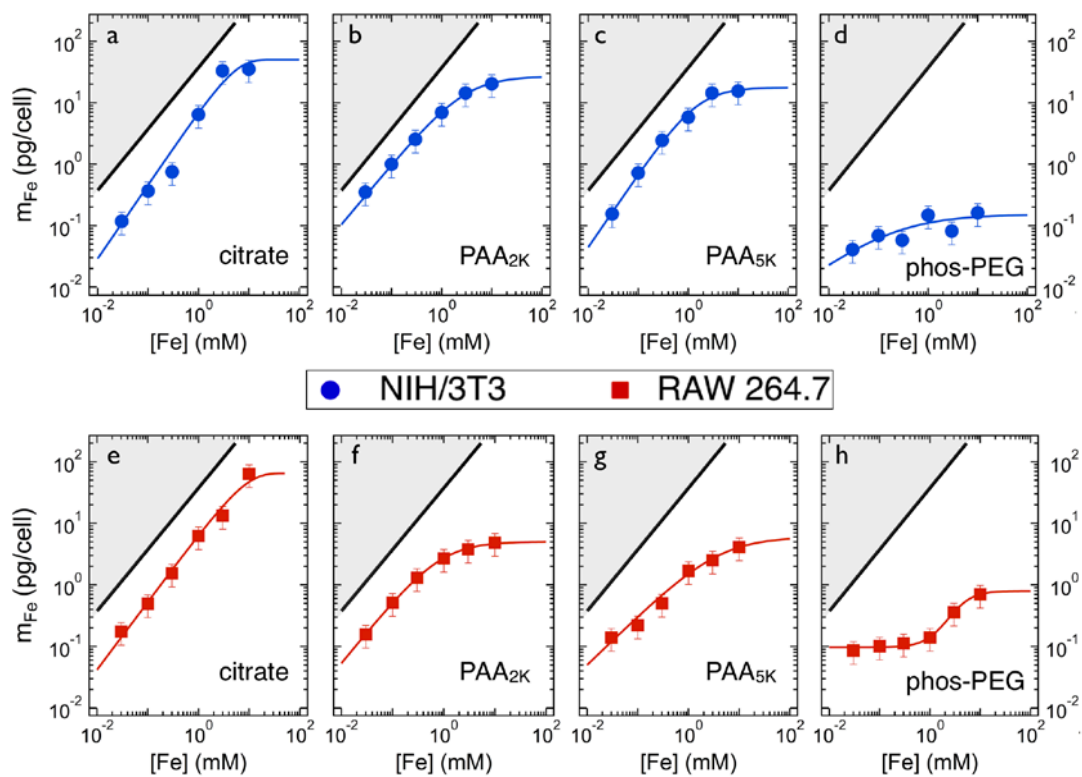

**Figure 6:** *Mass of iron atom per cell expressed in pg/cell as a function of the iron oxide concentration in the supernatant. $M_{Fe}$ is determined using the MILC protocol (**M**ass of metal **I**nternalized/Adsorbed by **L**iving **C**ells) described in Fig. 5.[47] Data are shown for NIH/3T3 fibroblasts (resp. RAW 264.7) and particles coated with citrate (a resp. e), poly(acrylic acid) 2100 g mol$^{-1}$ (b resp. f), poly(acrylic acid) 5100 g mol$^{-1}$ (c resp. g) and phosphonic acid PEG copolymers 12950 g mol$^{-1}$ (d resp. h). The straight thick lines in each figure depict the maximum amount of iron taken up by a single cell. It is calculated by dividing the mass of iron present in the supernatant by the number of cells in the assay. $3 \times 10^6$ fibroblasts exposed to a concentration of 1 mM corresponded to $m_{Fe} = 37$ pg/cell. With the units used, [Fe] = 1 mM corresponds to $c(\gamma\text{-}Fe_2O_3) = 8 \times 10^{-3}$ wt. % or 80 µg/ml.*

The straight thick lines in Fig. 6 depict the maximum amount of iron that can be taken up by a single cell at the treatment conditions. For citrate coated particles, $m_{Fe}$ increases linearly with [$Fe$] and leveled off above 10 mM (Figs. 6a and 6e). Here, the masses of internalized/adsorbed iron are high and in the range 50 – 70 pg/cell, independently of the cell lines. In the linear parts, it represents about 10 – 20% of the maximum value discussed previously. For particles coated with PAA$_{2K/5K}$ coating, the $m_{Fe}$-variations are similar but the saturation plateaus are at a lower level, between 5 and 20 pg/cell for NIH/3T3 and for RAW264.7 (Figs. 6b, 6c, 6f and 6g). The data with fibroblasts are in good agreement with those determined recently with smaller particles.[47] Note also that the carboxylic acid PEG coated particles exhibit the same behavior as those coated with PAA$_{2K/5K}$, the particles being slowly destabilized over time. Conversely, with phosphonic acid PEG copolymers, the uptake and adsorption levels are of the order of 0.1 pg/cell, *i.e.* 50 to 700 times below those found with citrate and poly(acrylic acid) (Figs. 6d and 6h). The monitoring of 3 types of behaviors in biofluids (from rapid precipitation to excellent dispersability) allows establishing a correlation between the stability of particles in cell culture media and the amount of adsorbed and internalized particles. Precipitating particles are prone to adsorb at the cellular membranes and to enter the cells, a process that is accelerated by sedimentation. For particles coated with polymers in contrast, sedimentation is less important. Significant differences are however observed between the charged $PAA_{2K/5K}$ and the neutral PEG coating. These differences were explained by the behavior of the polyelectrolyte brush in physiological solvent, which has the tendency to shrink at high ionic strength and to be less protective against aggregation.

Figs. 7a and 7b display the viability of NIH/3T3 and RAW264.7 cells treated with phosphonic acid PEG copolymers and with the 13.2 nm PEGylated particles. The toxicity assay was based on the WST1 protocol that assessed the mitochondrial activity. The experimental conditions were an incubation time of 24 h and an iron concentration between 0.03 mM and 10 mM. Viability experiments with phosphonic acid PEG copolymers (full symbols in Fig. 7) were performed at concentrations that were equivalent to those calculated from the amounts adsorbed on the particles. The results on the PEGylated copolymers and on the polymer coated particles showed no sign of toxicity, even at high doses, confirming the excellent biocompatibility of phosphonic acid PEG coating. The viability of particles coated with citrate and $PAA_{2K/5K}$ was also measured and it was in agreement with previous reports.[16,59]

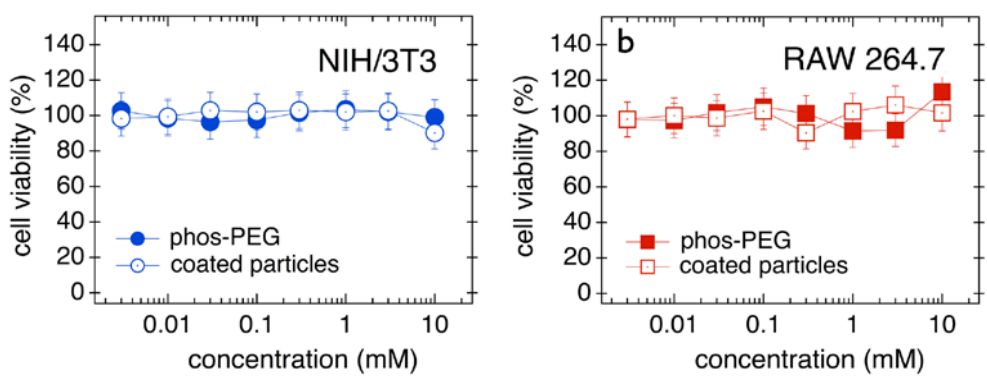

**Figure 7**: *Cell viability of NIH/3T3 and RAW264.7 treated with phosphonic acid PEG copolymers (full symbols) and with the 13.2 nm PEGylated particles (empty symbols) using the WST1 toxicity assay. The incubation time was 24 h. For the polymers, the concentration was equivalent to that calculated from the amounts adsorbed at the iron oxide interfaces, assuming a number of chains per particle $n_{ads}$ = 207.*

## IV - Conclusion

In this study, phosphonic acid poly(ethylene glycol) copolymers are synthesized and used as a coat for iron oxide particles. The copolymers put under scrutiny are consistent with 3 − 4 PEG chains of molecular weight 2000 g mol$^{-1}$ and 3 – 4 anchoring moieties grafted on the same methacrylate backbone. The interplay between the number of PEGylated chains, the nature of the anchoring groups and total molecular weight of the copolymers is optimized to allow a facile formulation of coated particles. With a polymer density of 0.5 nm$^{-2}$ at the iron oxide surface, 10 g of polymers (as obtained from a single synthesis) are sufficient to coat 40 g of iron oxide with diameter 13.2 nm. This later value may be compared to the mass of particles required in standard *in vitro* toxicity or *in vivo* biodistribution studies, which is of the order of the milligram. Time-resolved light scattering provides clear evidences that the coated particles are stable in biologically relevant conditions for months and are protective against proteins adsorption. The PEG coating efficacy is tested against different types of coating agents, including a copolymer with carboxylic acid groups *in lieu* of the phosphonic acid moieties. Carboxylic acid PEG copolymers show mitigated stability and protection against protein adsorption. The difference between the two polymers is interpreted in terms of binding affinity towards the iron oxide surface, which is assumed to be higher for phosphonic acid. Further work is being performed to clarify this issue. The present study confirms moreover previous findings: when proteins adsorbed at the particle surfaces and built a protein corona, they induce the agglomeration of the particles and *in fine* a destabilization of the dispersion. In this respect, the present study poses the question of the ubiquity of the protein corona and of its relevance in the context of applications related to nanomedicine. As the composition of the phosphonic acid PEG copolymers can be further implemented using stimuli responsive or fluorescent co-monomers, the strategy proposed here opens up new avenues for functionalizing inorganic surfaces.

## Supporting Information

The Supporting Information includes sections on the synthesis of poly(poly(ethylene glycol) methacrylate-*co*-methacrylic acid) (S1), on the reactivity of the monomers during synthesis (S2), on the characterization of the polymers by light scattering (S3) and by refractometry (S4), on the characterization of iron oxide nanoparticles by transmission electron microscopy (S5), by electron beam microdiffraction (S6), by vibrating sample magnetometry (S7) and zeta potential (S8). Section S9 displays the composition of the cell culture medium used and S10 shows the stability of the coated particles as a function of the salt concentration. This material is available free of charge via the Internet at http://pubs.acs.org

## Acknowledgements

We thank Armelle Baeza, Jean-Paul Chapel, Jérôme Fresnais, Fanny Mousseau, Emek Seyrek, Camille Vezignol for fruitful discussions. V.T. gratefully acknowledges the financial support for the stay in Paris from FIRB ITALNANONET, Project RBPR05JH2P_022 (MIUR, Rome, Italy). L.V. also thanks the CNPq (Conselho Nacional de Desenvolvimento Científico e Tecnológico) in Brazil for postdoctoral fellowship. The Laboratoire Physico-chimie des Electrolytes, Colloïdes et Sciences Analytiques (UMR Université Pierre et Marie Curie-CNRS n° 7612) is acknowledged for providing us with the magnetic nanoparticles. ANR (Agence Nationale de la Recherche) and CGI (Commissariat à l'Investissement d'Avenir) are gratefully acknowledged for their financial support of this work through Labex SEAM (Science and Engineering for Advanced Materials and devices) ANR 11 LABX 086, ANR 11 IDEX 05 02.